\documentclass[pdflatex,sn-mathphys-num]{sn-jnl}

\usepackage{graphicx}%
\usepackage{multirow}%
\usepackage{amsmath,amssymb,amsfonts}%
\usepackage{amsthm}%
\usepackage{mathrsfs}%
\usepackage[title]{appendix}%
\usepackage{xcolor}%
\usepackage{textcomp}%
\usepackage{manyfoot}%
\usepackage{booktabs}%
\usepackage{algorithm}%
\usepackage{algorithmicx}%
\usepackage{algpseudocode}%
\usepackage{listings}%

\theoremstyle{thmstyleone}%
\theoremstyle{thmstyletwo}%

\theoremstyle{thmstylethree}%

\begin{document}

\title[Article Title]{Marginal Productivity Theory versus \\
	the Labor Theory of Property: \\An analysis using vectorial marginal products}

\author{\fnm{David} \sur{Ellerman} \email{david@ellerman.org}}

\affil{ \orgdiv{Faculty of Social Sciences}, \orgaddress{University of Ljubljana}, \city{Ljubljana}, \state{Slovenia}, \country{orcid.org/0000-0002-5718-618X}}


\abstract{\noindent Neoclassical economic theory presents marginal productivity (MP)
	theory using the scalar notion of marginal products, and takes pains,
	implicitly or explicitly, to show that competitive equilibrium satisfies the
	supposedly ethical principle: ``To each what he and the
	instruments he owns produces.'' This paper shows that MP
	theory can also be formulated in a mathematically equivalent way using
	vectorial marginal products--which however conflicts with the above-mentioned
	``distributive shares'' picture. Vectorial MP
	theory also facilitates the presentation of modern treatment of the labor theory of property which on
	the descriptive side is based on the fact that, contrary to the distributive
	shares picture, one legal party gets the production vector consisting of 100
	percent of the liabilities for the used-up inputs and 100 percent of the
	produced outputs in a productive opportunity. On the normative side, the labor theory of property is just the application of the usual juridical norm of
	imputation to the question of property appropriation.}




\maketitle
\tableofcontents
\section{Introduction}\label{sec1}

When many economists and philosophers consider the principle of people getting
the fruits of their labor, they will usually interpret it in terms of marginal
productivity. John Rawls is a good example.

\begin{quotation}
	\noindent Accepting the marginal productivity theory of distribution, each
	factor of production receives an income according to how much it adds to
	output (assuming private property in the means of production). In this sense,
	a worker is paid the full value of the results of his labor, no more and no
	less. Offhand this strikes us as fair. It appeals to a traditional idea of the
	natural right of property in the fruits of our labor. Therefore to some
	writers the precept of contribution has seemed satisfactory as a principle of
	justice. \cite[p. 271]{rawls:justice}
\end{quotation}

\noindent The neoclassical claim is that under the conditions of competitive
equilibrium, each unit of labor ``gets what it
produces.'' Indeed, Milton Friedman calls it the
``capitalist ethic'' \cite[p. 164]
{friedman:capandfreedom}:

\begin{quotation}
	\noindent The ethical principle that would directly justify the distribution
	of income in a free market society is, ``To each what he and
	the instruments he owns produces.'' \cite[pp. 161-2]
	{friedman:capandfreedom}
\end{quotation}

Well-meaning liberal, progressive, and even heterodox economists emphasize
that the actual economy may be neither competitive nor in equilibrium, that
fixed proportions in production (Sraffa) raise enormous difficulties in
measuring the marginal product of each factor of production, and that there
may be nothing just in the given distribution of wealth. But they raise no
objection in principle to the interpretation of marginal productivity theory
as giving people ``what they produce''; they
only fuss about its applicability in practice as well as about the prior
personal distribution of factor ownership.\footnote{In addition to John Rawls
	\cite{rawls:justice}, see Chapter 6, ``To each according to
	his contribution'' in Steve Keen \cite{keen:debunk} or
	Appendix A in Lester Thurow \cite{thurow:inequality}.}

But is this ``question of distribution'' even
the most important question? There is another more fundamental question that
is usually passed over without discussion, and that is the ``predistributive'' \cite{ellerman:challengelpt} question of:
``Who is to be the firm in the first place?'',
e.g., ``capital,'' labor, the government, an
entrepreneur, and so forth. In the technical treatments of neoclassical
economics, a productive opportunity is technologically described by a set of
production vectors where each vector's negative components represent the
used-up inputs (liabilities leading to expenses) and positive components
represent the produced outputs (assets leading to revenues). This production
vector is also called a ``production possibility
vector'' \cite[p. 267]{arrow-debreu:gemodel}, an
``activity vector'' \cite[p. 59]%
{arrow-hahn:gca}, a ``production'' \cite[p.
38]{debreu:theoryofvalue}, or an ``input-output
vector'' \cite[p. 27]{quirk-saposnik}. Hence the
predistributive question could be formulated as: ``Who is to
get the (liabilities and assets in the) production vector of a productive
opportunity?''.

This question of who is to be the firm, in the sense of the ``appropriator of the production vector'', is obviously not the
same as the question of distribution or distributive shares. Hence before
considering the problems in the interpretation and formulation of marginal
productivity theory, we have to first consider the predistributive question.

\section{Who gets the production vector?}

One can give an institution-free description of a productive opportunity in
concrete detail without specifying ``who is the
firm'', e.g., such and such people are using these types of
machines along with certain other inputs to produce a list of products. In the
rare occasions when the question of which legal party owes the
input-liabilities (negative components in the production vector) and owns the
produced-assets (positive components), then the answer is usually:
``The owner of the firm--of course!'' This
answer comes in many forms. Economists might try to directly connect their
mathematical formalism to the legal system. For instance, the production set
may be presented as the ``production-opportunity
locus'' \cite[p. 124]{hirsh:invest}, and then economists
might speak of the ``owners'' of these
technical possibilities, e.g., the ``owners of the productive
opportunity'' \cite[p. 125]{hirsh:invest}. Other economists
try similarly to address the predistributive question by just postulating a
legal institution corresponding to their analytical notion of a production function.

\begin{quotation}
	\noindent Indeed, how should we define an ``entrepreneur''? It seems that he is
	not just a capital-owner, or one who has the right of disposal over capital.
	He is not simply a manager, because as such he could be counted among the
	employees. He could be said to be the party who gets the net profits. But for
	what? This is a very old subject of debate. Perhaps the best way out is to
	define the entrepreneur as the ``owner of a production function''. In this way
	he has some sort of exclusive right. Nobody else can use his production
	function. \cite[pp. 209-10]{haavelmo:capital}
\end{quotation}

\noindent But one hardly needs to scan the legal literature to find there is
no such legal institution as the ownership of a production set, production
function, production-opportunity locus, activity vector, or the like.

The widely acclaimed Arrow-Debreu model attempted to show that there could be
a competitive equilibrium with:

\begin{quotation}
	\noindent a category of pure profits which are distributed to the
	owners of the firm; it is not assumed that the owners are necessarily the
	entrepreneurs or managers. ... In the McKenzie model, on the other hand, the
	firm makes no pure profits (since it operates at constant returns); the
	equivalent of profits appears in the form of payments for the use of
	entrepreneurial resources, but there is no residual category of owners who
	receive profits without rendering either capital or entrepreneurial services.
	\cite[p. 70]{arrow:firmingetheory}
\end{quotation}

\noindent The first question any economist familiar with the logic of
perfectly idealized competitive markets would (or, at least, should) ask is:
``why wouldn't some one offer a slightly higher price for the
inputs and accept a slightly lower profit, but thereby take over the
productive opportunity?'' The `answer' that Arrow-Debreu
offered in their formalism was that production sets had ``owners'' represented by shares with given owners. Leaving
aside the crude attempt to marry analytical concepts and legal institutions,
Arrow and Debreu clearly had in mind the notion of a corporation with
shareholders.\footnote{See Chapter 4 in \cite{ellerman:juris-book} for a
	fuller discussion of the analytical flaws in the Arrow-Debreu attempt to show
	that there could be an equilibrium in competitive markets with positive pure
	profits.}

This brings us to the other common answer to the predistributive question:
``The owners of the corporation''--as if the
ownership of the production vector resulting from a productive opportunity
undertaken using the capital assets of a corporation was part and parcel of
the ``ownership of a corporation.'' But once
again, economists who understand markets know that the capital assets of a
corporation can be rented out to some other party undertaking a productive
opportunity using those assets--rather than the corporation renting or
purchasing a complementary set of inputs and undertaking the opportunity
itself. 

It is routine for capital assets like buildings and machinery to be leased
out. There is even a historical example of leasing out a whole factory. In the
early 1950s, the Studebaker-Packard Corporation had the Packard bodies
produced by the Briggs Manufacturing Company in their Conner Avenue plant in
Detroit. When the founder died, all twelve of the U.S. Briggs plants were sold
to the Chrysler Corporation in 1953. ``The Conner Ave. plant
that had been building all of Packard's bodies was leased to Packard to avoid
any conflict of interest.'' (Theobald 2004) Then the
Studebaker-Packard Corporation would hold the management rights and production
vector rights for the operation of the factory \textit{owned} by the Chrysler Corporation.

The logical point to understand is that ``who gets the
production vector?'' is determined by the pattern of market
contracts, e.g., who hires, rents, leases what or who, and that question is
not answered by the given distribution of share ownership of corporations.
Hence all the economists' answers to the predistributive question like
``the owners of the firm, corporation, or
company'' ignore the role of market activity which determines
who gets the production vector. Assets are owned by corporations but patterns
of customer or supplier market contracts are not `owned' by corporations.

A simple example of this misunderstanding in the standard textbooks is the
``circular flow diagram'' with input
suppliers arrayed on one side of the picture and ``firms'' arrayed on the other side with factor markets in
between. But we have just seen that it is only the pattern of the factor
market contracts themselves that determines which legal parties end up on the
``firm'' side of the market, e.g., the
Chrysler corporation ended up as an ``input
supplier'' side rather than as a ``firm'' or ``input demander'' in the productive opportunity undertaken by another car company leasing its factory.

Hence there are two basic questions: the question about the distributive
shares usually considered by economists and the prior predistributive question
of ``who is to be the firm in the first
place?''--where by ``firm'' is meant the legal party that gets the production vector from a
technologically possible production set. Both questions have a descriptive
version and a normative version. For instance, we have shown that the
descriptive predistributive question is not answered by the pattern of given
ownership of labor, capital, corporate shares or the like; it is answered by
the pattern of market contracts. There is no small irony when conventional
economists invent imaginary legal notions like the ``ownership
of the production function'' to explain what is in fact
explained by the direction of market contracts--by who hires what or whom.

What about the two normative questions? The answers given to both the
distributive and predistributive question use some notion of ``responsibility.'' The underlying norm, often unspoken, is
that people should ``get'' whatever they are
responsible for. For instance, there is a long tradition to say that factor
payments according to marginal productivity would satisfy this principle of
justice. But his notion sees both human and non-human factors as being
``economically responsible'' which is clearly
not the usual legal or moral notion of responsible agency which can only apply
to persons.

Hence there is another point of view that if we analyze production from the
point of view of the de facto responsibility where only persons qualify, then
only the persons involved in a productive opportunity could be de facto
responsible for \textit{both} using up the inputs and producing the outputs so
by applying the usual juridical imputation principle, they should get that
production vector. Hence the usual notion of legal responsibility is used to
argue, in the modern version of the labor or natural rights theory of
property, that the people who work in a productive opportunity should
constitute the legal party that owes the production liabilities (the negative
fruits of their labor) and should own the produced assets (the positive fruits
of their labor). This argument answers the predistributive question with:
``Labor should get the production vector'' which asserts in the older language of the 19th century ``Labour's Right to the Whole Product'' \cite{menger:wholeproduct} (where in modern terms, ``whole
product'' = production vector).

With these two very different questions and answers clearly understood, we can
circle back and analyze the usual treatment of marginal productivity theory.

\section{Analysis of marginal productivity theory}

The modern treatment of the labor theory of property is carried out in more
detail elsewhere (\cite{ellerman:pandc}, \cite{ellerman:intro-lpt},
\cite{ellerman:juris-book}) where it is argued that employer-employee system
does not satisfy, even in principle, the norm of people getting the fruits of
their labor. The orthodox view of marginal productivity theory is flawed on
several counts which might be called:

\begin{itemize}
	\item the metaphor,
	
	\item the mistake, and
	
	\item the miracle.
\end{itemize}

\subsection{The Metaphor: Only persons have responsible agency}

The first and foremost problem is the neglect of the difference between
responsible human actions and the non-responsible but causally efficacious
(i.e., productive) services of things like a wrench, machine, or truck. This
fundamental distinction between persons and things in terms of imputation is
part of standard jurisprudence.

\begin{quotation}
	\noindent A \textit{person} is the subject whose actions are susceptible to imputation.
	\ldots\ A \textit{thing} is something that is not susceptible to imputation.
	\cite[pp. 24-25]{kant:metamorals}
\end{quotation}

The distinction is hardly new or subtle except perhaps in orthodox economics
where both human actions and the services of things are often treated simply
as causally effective productive services. Frank Knight was perhaps the most
philosophically sophisticated defender of orthodox economics and the current
economic system.

\begin{quotation}
	\noindent We have insisted that the word ``produce'' in the sense of the specific [i.e., marginal]
	productivity theory of distribution, is used in precisely the same way as the
	word ``cause'' in scientific discourse in
	general. \cite[p. 178]{knight:risk}
	
	\noindent For ``labor'' we should now say
	``productive resources.'' \cite[p.
	8]{knight:history}
\end{quotation}

Conventional economics, across the whole spectrum from Knight and the Chicago
school on the right to progressive economists like Joseph Stiglitz, Thomas
Piketty, James Galbraith, Amartya Sen, Edmund Phelps, or Sam Bowles on the
left, frame the most basic problem as the problem of distribution--and seen
unable to frame the predistributive question of ``Who is to be
the firm'' in the first place.

\begin{quotation}
	\noindent Goods are typically produced by the co-operation of various kinds of
	productive services, and the special problem of distribution, in modern terms,
	is that of the division of this joint product among the different kinds of
	co-operating productive services and agents. \cite[p. 21]{knight:history}
\end{quotation}

If economists from the right or left were on jury duty for a murder trial,
they would probably drop their learned ignorance of difference between the
responsible actions of persons and the causally efficacious services of
things. They would probably not wonder---or, at least, not out loud---how to
effect the ``division'' of the joint
responsibility ``among the different kinds of co-operating
productive services and agents.'' They might even understand
that the responsibility for the murder is imputed back through any gun or
other weapon to the person using those instruments.

One has to go back to before the distributive shares metaphor conquered neoclassical economics (like the ``Holy Inquisition conquered Spain'' as Keynes might say) to find a non-metaphorical statement about actual legal
or moral imputation. In the early days of marginal productivity theory (1889),
the legally-trained Austrian economist, Friedrich von Wieser, pointed out the
simple facts.

\begin{quotation}
	The judge ... who, in his narrowly-defined task, is only concerned with the
	legal imputation, confines himself to the discovery of the legally responsible
	factor, --that person, in fact, who is threatened with the legal punishment.
	On him will rightly be laid the whole burden of the consequences, although he
	could never by himself alone--without instruments and all the other
	conditions--have committed the crime. The imputation takes for granted
	physical causality. \cite[p.76]{wieser:value}
\end{quotation}

\subsection{The Mistake: One party gets the whole production vector}

In terms of actual \textit{property rights}, the shares in the product are not
actually imputed or assigned to the various factor suppliers. For instance,
employees are paid for their labor services even if, by some accident, no
product is produced. When the product is produced, they are not paid for their
product since they don't own the product in the first place in order to sell
it (since they didn't pay the costs to produce it). One legal party appropriates the ``whole
product'' of a firm, 100 percent of the output assets and 100
percent of the input liabilities. Modern economists display a learned
ignorance of this simple legal fact when thus use the ``distributive shares'' framing of the question. It seems one
has to also go back to before the development of MP theory to find a statement
of the simple legal fact that one party covers all the liabilities for
``both instruments of production'' capital
and labor, and owns the whole of the produce. James Mill was quite forthright.

\begin{quotation}
	\noindent The owner of the slave purchases, at once, the whole of the labour,
	which the man can ever perform: he, who pays wages, purchases only so much of
	a man's labour as he can perform in a day, or any other stipulated time. Being
	equally, however, the owner of the labour, so purchased, as the owner of the
	slave is of that of the slave, the produce, which is the result of this
	labour, combined with his capital, is all equally his own. In the state of
	society, in which we at present exist, it is in these circumstances that
	almost all production is effected: the capitalist is the owner of both
	instruments of production: and the whole of the produce is his. \cite[Chapter
	I, section II]{milljames:elements}
\end{quotation}

Or one can venture outside of `economics' proper to find statements of the
actual facts about ownership of the product. For instance, an economic
sociologist put it simply over a century ago.

\begin{quotation}
	\noindent Under the factory system, the factory, raw materials, and finished
	product belong to the capitalist. The laborer at no time owns any part of what
	is passing through his hands or under his eye. Never can he say,
	``This product, when finished, will be mine, and my rewards
	will depend on how successfully I can dispose of it.'' There
	is much theoretic discussion to the ``right of labor to the
	whole product'' and much querying as to how much of the
	product belongs to the laborer. These questions never bother the manufacturer
	or his employee. They both know that, in actual fact, all of the product
	belongs to the capitalist, and none to the laborer. The latter has sold his
	labor, and has a right to the stipulated payment therefor. His claims stop
	there. He has no more ground for assuming a part ownership in the product than
	has the man who sold the raw materials, or the land on which the factory
	stands. \cite[pp. 65-6]{fairchild:appliedsoc}
\end{quotation}

One may search in vain through the entire corpus of modern economics to find
such a plain statement of the ``actual fact'' that the employer owns 100\% of the produced outputs and owes 100\% of the
liabilities for the used up inputs---with the employees having 0\% of both.
The de facto responsible actions of the employees are simply one of the liabilities.

\subsection{The Miracle: Virgin birth of marginal products}

There is still another flaw in the orthodox treatment of MP theory. The
ideological baggage being carried by MP theory forces it to be presented in a
factually implausible way, i.e., the ``virgin
birth'' of marginal products. This factually implausible part
of the orthodox view is the picture of a unit of a factor as ``immaculately'' producing its marginal product without using
up any other inputs. When another unit of an input $x$ is used, then by
considering a hypothetical shift to a slightly more $x$-intensive production
process, more output can be produced using the same amount of the other
factors, and that extra output is the ``marginal
product'' of that $x$ factor. But cost-minimizing firms would
not make such a hypothetical shift when increasing factor usage; they would
expand along the least-cost expansion path. More of other factors must then be
used, so the additional factors and product would be given by a vector, a
\textit{vectorial marginal product}. The simple ``distributive
shares of the product'' framing collapses since the vectorial
marginal product includes both a ``share'' of
the output but also an increase in using up other inputs to expand along the
least-cost expansion path.

In this essay, we give the mathematically equivalent vectorial presentation of
MP theory, which is based on the more plausible picture that a unit of labor
can only produce more of the outputs by using up more of the other inputs at
minimum cost. The ``problem'' with this
version of MP theory is that it does not lend itself to the ideologically
appealing picture of each unit of a factor as ``producing its
marginal product.'' Thus we have a central example about how
the ideological baggage being towed by orthodox economics affects even the
mathematical presentation of the standard theories.

\section{The Conventional Picture of Scalar Marginal Products}

Marginal productivity (MP) theory has always played a larger importance in
orthodox economics than could be justified by its purely analytical role (as a theory of factor demand). This
is because MP theory is conventionally interpreted as showing that, in
competitive equilibrium, ``each factor gets what it is
responsible for producing.'' The marginal unit of a factor is
seen as producing the marginal product of that factor, and each unit could be
taken as the marginal unit, so each unit ``produces its
marginal product.'' Consider the \textit{marginal physical
	product of labor} $MP_{L}$. In competitive equilibrium, the value of the
marginal product of labor $pMP_{L}$ (where $p$ is the unit price of the
output) is equal to $w$ (the unit price of labor):

\begin{center}
	\qquad$pMP_{L}=w$
	
	\qquad''Value of what a unit produces'' =
	``Value received by a unit of the factor.''
\end{center}

\noindent As Knight put it:

\begin{quotation}
	\noindent Since in a free market there can be but one price on any commodity,
	a general wage rate must result from this competitive bidding. The rate
	established may be described as the socially or competitively anticipated
	value of the laborer's product, using the term ``product'' in the sense of specific contribution, .... \cite[p.
	274]{knight:risk}
\end{quotation}

Our purpose is to highlight an internal incoherence in the conventional
treatment, to show how this difficulty can be overcome in a mathematically
equivalent reformulation of MP theory, and to note how this reformulation
accommodates a rather different interpretation of the theory.

The problem (or internal incoherence) in the usual treatment is simply that a
unit of a factor cannot immaculately produce its marginal product out of
nothing (the ``miracle''). The factor must
simultaneously use some of the other factors. If the marginal product of one
person-year in a tractor factory is one tractor, how can a tractor be produced
without using steel, rubber, energy, and so forth? But when that
concurrent factor usage is taken into account (``priced
out''), then the usual equations must be significantly
reformulated. A vectorial notion of the marginal product, the
``vector marginal product,'' must be used in
place of the conventional scalar marginal product.

Before turning to the vectorial treatment of marginal products, we must remove
the seeming paradox in the immaculate scalar treatment. When we increase the
labor in a tractor factory to produce more tractors, we will also have to
increase the steel, rubber, energy, and other inputs necessary to produce
tractors. That would spoil the attempt to take the increase in tractor output
as the result of solely the increase in labor. But the so-called
``marginal product of labor'' is the result
of a somewhat different hypothetical or conjectural change in production. It
is assumed that factors are substitutable. To arrive at the ``marginal product of labor'' we must consider two changes: an
increase in labor and a shift to a slightly more labor-intensive production
technique so that the increased labor can be used together with exactly the
same total amounts of the other factors. Since (following the hypothetical
production shift) the other factors are used in the same total amounts, the
extra output is then viewed as the ``product'' of the extra unit of labor, as if the extra product was miraculously
produced \textit{ex nihilo} by the extra unit of labor.

There is one other point that might be mentioned. Since the usual
``story'' represents each factor as producing
a share of the product (incurring no other costs) and getting the value of
that share, one might wonder if there is a dual metaphor about the
distribution of the costs. Indeed, there is. The metaphorical picture of each
input-supplier as ``producing'' a share of
the outputs through the inputs supplied, dualizes to the picture of each
output-demander as using up a share of the inputs consumed in producing the
unit of output demanded. The value of those used-up inputs at the margin is
the marginal cost $MC$ so the dual part of the ``capitalist
ethic'' is that the output-demander should owe for those
liabilities and thus pay the price $p=MC$. But in terms of non-metaphorical
property assets and liabilities, the input-suppliers do not own shares of the
output assets, and the output-demanders do not owe for a share of the total
input liabilities.\footnote{One might wonder if those who (metaphorically) see
	``capital suppliers'' as ``participating'' in a company to produce the outputs--would
	also see the customers as ``participating'' in the company to use up the inputs. The mathematics is symmetric,
	particularly for production vectors with many inputs and many outputs.}

The two metaphors cancel out. There is in fact one legal party (sometimes
called the \textit{residual claimant} or simply the ``firm'') which stands between the input suppliers and output
demanders, and, as already noted, that legal party appropriates (i.e., owes)
100 percent of the input liabilities and (i.e., owns) 100 percent of the
output assets, i.e., legally appropriates the \textit{production vector} or,
in alternative terminology, the \textit{whole product vector}.

The fundamental question about production is \textit{not} about distributive
shares in asset or liability values, but the prior \textit{predistributive}
question: ``Who is to be the firm in the first place: Capital,
Labor, or the State?''\cite{ellerman:challengelpt}, i.e., who
is to appropriate the whole production vector. That is the question addressed
by a theory of property (\cite{ellerman:pandc}, \cite{ellerman:intro-lpt},
\cite{ellerman:juris-book}), not a theory of value dealing with input or
output prices.

\section{Symmetry Restored: The Pluses and Minuses of Production}

Nothing is produced \textit{ex nihilo}. Labor cannot produce tractors without
actually using other inputs. Production needs to be reconceptualized in an
algebraically symmetric manner. That is, there are both positive results
(produced outputs) and negative results (used-up inputs), and they can be
considered symmetrically in a vectorial form.

For a nontechnical presentation, let $Q=f(K,L)$ be a production function with
$p$, $r$, and $w$ as the unit prices of the outputs $Q$, the capital services
$K$, and the labor services $L$ respectively. The outputs $Q$ are the positive
product of production but there is also a negative product, namely the used-up
capital and labor services $K$ and $L$. Lists or vectors with three components
can be used with the outputs, capital services, and labor services listed in
that order. The \textit{positive product} would be represented as $(Q,0,0)$.
The \textit{negative product} signifying the used-up or consumed inputs could
be represented as $(0,-K,-L)$. The comprehensive and algebraically symmetric
notion of the product is obtained as the (component-wise) sum of the positive
and negative products. It might be called the \textit{whole product }(or
production vector) [where symbols for vectors are in bold].\footnote{This
	vectorial treatment of the whole production vector \cite{varian:micro} is now
	standard in the literature of mathematical economics, but without singling out
	any responsible factor.}

\begin{center}
	$\mathbf{WP}=(Q,-K,-L)=(Q,0,0)+(0,-K,-L)$
	
	Whole Product = Positive Product + Negative Product
\end{center}

The unit prices can also be arranged in a vector, the \textit{price vector}
$\mathbf{P}=(p,r,w)$. The (dot) product of a price vector times a quantity
vector (such as the whole product vector) is the sum of the component-wise
products of prices times quantities. That sum is the value of the quantity vector.

\begin{center}
	$\mathbf{P\cdot}(Q,0,0)=(p,r,w)\cdot(Q,0,0)=pQ$
	
	Value of Positive Product = Revenue
	
	$\mathbf{P}\cdot(0,-K,-L)=(p,r,w)\cdot(0,-K,-L)=-(rK+wL)$
	
	Value of Negative Product = Expenses
	
	$\mathbf{P}\cdot(Q,-K,-L)=(p,r,w)\cdot(Q,-K,-L)=pQ-(rK+wL)$
	
	Value of Whole Product = Profit
\end{center}

\section{Marginal Whole Products}

\begin{figure}[h]
	\centering
	\includegraphics[width=0.7\linewidth]{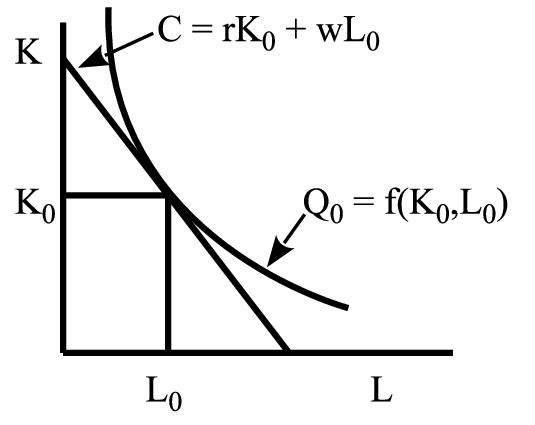}
	\caption{Minimum Cost to Produce Quantity $Q_{0}$}
	\label{fig:fig1-mincost}
\end{figure}

The alternative presentation of MP theory uses the marginal version of the
whole product vector, which we will call the ``marginal whole
product.'' The full mathematical development is given in the
Appendix. Here we develop a heuristic discrete treatment. Given the input
prices and a given level of output $Q_{0}$, there are input levels $K_{0}$ and
$L_{0}$ that produce $Q_{0}$ at minimum cost $C_{0}=rK_{0}+wL_{0}$ as illustrated in Figure \ref{fig:fig1-mincost}.

\begin{figure}[H]
	\centering
	\includegraphics[width=0.7\linewidth]{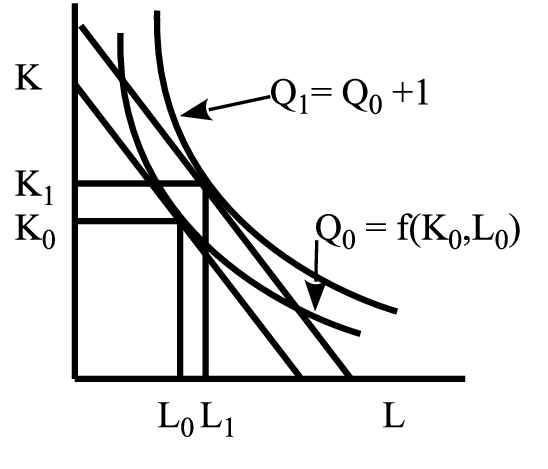}
	\caption{New Levels of $K$ and $L$ to Produce $Q_{1}=Q_{0}+1$}
	\label{fig:fig2-change}
\end{figure}

For an increase of one unit of output to $Q_{1}=Q_{0}+1$, there will be new
levels of $K_{1}$ and $L_{1}$ necessary to produce $Q_{1}$ at minimum cost, as shown in Figure \ref{fig:fig2-change}.

Let $\Delta K=K_{1}-K_{0}$ and $\Delta L=L_{1}-L_{0}$ be the marginal
increases in the amounts of capital and labor services that are necessary to
produce the increase in output $\Delta Q=Q_{1}-Q_{0}=1$. The minimum cost of
producing $Q_{1}$ is $C_{1}=rK_{1}+wL_{1}$. Since $\Delta Q=1$, the marginal
cost is:

\begin{center}
	$MC=\Delta C/\Delta Q=\Delta C=C_{1}-C_{0}=rK_{1}+wL_{1}-(rK_{0}+wL_{0})$.
\end{center}

The marginal version of the whole product is the \textit{marginal whole
	product} which has unit output and minus the inputs necessary to produce one
more unit of output at minimum cost.

\begin{center}
	$\mathbf{MWP}=(1,-\Delta K,-\Delta L)$
	
	Marginal Whole Product
\end{center}

The value of the marginal whole product is the \textit{marginal profit}, the
difference between price and marginal cost.

\begin{center}
	$\mathbf{P}\cdot\mathbf{MWP}=(p,r,w)\cdot(1,-\Delta K,-\Delta L)=p-[(rK_{1}%
	+wL_{1})-(rK_{0}+wL_{0})]=p-MC$.
	
	Value of Marginal Whole Product is Marginal Profit
\end{center}

If the marginal profit was positive at a given level of output, then profits
could be increased by increasing the level of output. If the marginal profit
was negative, then profits would increase by decreasing the level of output.
Thus if profits are at a maximum, then the marginal profit must be zero. This
is the usual result that $p=MC$ if profits are at a maximum.

\section{Asymmetry Between Responsible and Non-Responsible Factors}

Part of the poetic charm of the conventional presentation of MP theory was
that it allowed each factor to be pictured as active---as being
``responsible'' for producing its own
marginal product. But we have noted the technical absurdity of, say, labor
producing tractors out of nothing else. Labor must use up steel, rubber, and
other inputs to produce tractors. But if that is accepted, then it is
implausible to turn around and pretend that another factor is also
active---that steel uses up labor, rubber, and other factors to produce
tractors. Hence neoclassical economics uses the usual picture of each factor
as ``producing its marginal product'' without
using other factors--which then allows it to invoke, as Milton Friedman put
it: ``the ethical proposition that an individual deserves what
is produced by the resources he owns'' \cite[p.
199]{friedman:pricetheory}.

MP theory, as an analytical economic theory, does not provide any distinction
between responsible or non-responsible factors. Those notions must be imported
from jurisprudence \cite{ellerman:juris-book}. No amount of staring at partial
derivatives will reveal the difference between responsible and non-responsible
factors. ``Responsibility'' is a
legal-jurisprudential notion. Neoclassical theory sometimes uses poetic
license and the pathetic fallacy to represent all the factors as being
responsible and cooperating together to produce the outputs. For instance, as
Paul Samuelson put it: ``Together, the man and shovel can dig
my cellar'' or ``land and labor together
produce the corn harvest'' \cite[pp. 536-7]{samuelson:economics}. But poetry and pathetic fallacy aside, a man uses a shovel to dig a cellar
and people use land (and other inputs) to produce the corn harvest. Only human
actions can be responsible for anything. For example, the tools of the
burglary trade certainly have a causal efficacy (``productivity''), but only the burglar can be charged with
responsibility for the crime. The responsibility is imputed back through the
tools (as ``responsibility conduits'') to the
human user.

Friedrich von Wieser, who was trained in jurisprudence, introduced the notion
of imputation into economics to metaphorically talk about the
``responsible agency'' of all the agents of
production. But as noted above, he was quite clear that for the
non-metaphorical notions of legal or moral imputation, only persons could be responsible.

\begin{quotation}
	\noindent If it is the moral imputation that is in question, then certainly no
	one but the labourer could be named. Land and capital have no merit that they
	bring forth fruit; they are dead tools in the hand of man; and the man is
	responsible for the use he makes of them. \cite[p. 79]{wieser:value}
	
	\noindent As soon as the judge has established the causal nexus and the
	presumption of sanity, he is bound to attribute the entire result to the
	accused. This is true even though he may know very well that the accused could
	never have accomplished it alone without instruments and without the peculiar
	contributing circumstances. \cite[p. 115]{wieser:social}
\end{quotation}

\noindent This admission about non-metaphorical legal and moral imputation was
only made in the early days of establishing the \textit{metaphorical}
reinterpretation of imputation, ``economic
responsibility,'' to be used in the presentation of MP theory.

\begin{quotation}
	\noindent In the division of the return from production, we have to deal
	similarly ... with an imputation, -- save that it is from the economic, not
	the judicial point of view. \cite[p. 76]{wieser:value} THE ECONOMICALLY
	RESPONSIBLE FACTORS \cite[header on p. 77]{wieser:value}
\end{quotation}

\noindent The modern texts just present this metaphorical ``economic'' imputation as if it was the only imputation.

Property theory, on the other hand, should deal with \textit{non-metaphorical}
legal and moral imputation precisely from ``the judicial point
of view.'' Property theory \cite{ellerman:intro-lpt} on its
normative side was traditionally called the ``labor theory of
property''\footnote{See, for instance, Thomas Hodgskin
	\cite{hodgskin:property} and Foxwell's introduction \cite{foxwell:intro} to
	the book by Carl Menger's legally-trained brother, Anton Menger
	\cite{menger:wholeproduct}.} but in its modern form, it is just the usual
\textit{juridical principle of imputation}: Assign legal rights and
liabilities to the de facto responsible agents (\cite{ellerman:pandc},
\cite{ellerman:juris-book}). For instance, only the persons (including
managers) working in a productive opportunity are de facto responsible for
using up the inputs in the process of producing the outputs, and thus they
should jointly owe those legal liabilities (used-up inputs = negative fruits
of their labor) and have the legal ownership of those assets (produced outputs
= positive fruits of their labor), i.e., they should jointly appropriate the
\textit{whole product of labor} (or \textit{labor product}) which can be seen
as the whole product $\mathbf{WP}=\left(  Q,-K,-L\right)  $ plus the labor
services $\left(  0,0,L\right)  $ viewed as a commodity (that they
``create'' and ``use
up'').

\begin{center}
	$\mathbf{WP}_{\mathbf{L}}=\left(  Q,-K,0\right)  =\mathbf{WP}+(0,0,L)$
	
	Whole product of labor.
\end{center}

Since MP theory does not, by itself, provide any concept of ``responsible'' factors, any factor or factors could be taken
as the responsible factors for analytical purposes. In the mathematical
treatment given below, the factors $x_{1},...,x_{n}$ will not be identified
(as capital, labor, etc.), and we will arbitrarily take the first factor as
being responsible. In our nontechnical presentation where the factors are
identified, labor will taken as the responsible factor (but the
\textit{formalism} would be the same, \textit{mutatis mutandis}, for any other choice).

As the responsible factor produces the outputs (produces the positive
product), it must also use up the inputs (produce the negative product). We
must calculate the positive and negative product of the marginal unit of the
responsible factor, labor. We will call the vector of positive and negative
marginal results of labor, the ``marginal whole product of
labor.'' The value of marginal whole product of labor is then
compared with the opportunity cost of labor (the wage $w$ in the model).

\begin{figure}[h]
	\centering
	\includegraphics[width=0.7\linewidth]{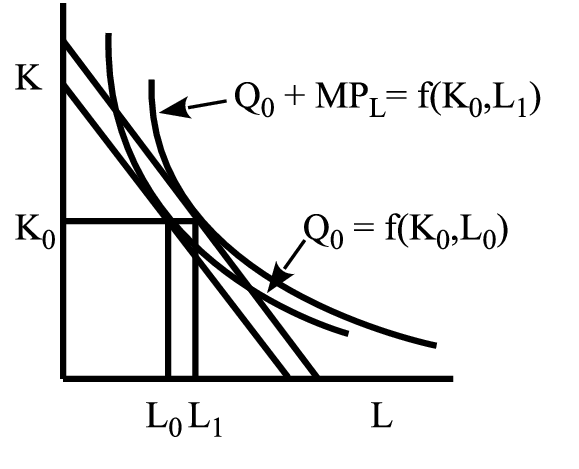}
	\caption{$MP_{L}$ as the increase in $Q$ from $\Delta L$ keeping capital
		constant at $K_{0}$}
	\label{fig:fig3-mpl}
\end{figure}

The marginal quantities $\Delta Q=1$, $\Delta K$, and $\Delta L$ that appear
in the marginal whole product can be used to form the ratios $\Delta Q/\Delta
K$ and $\Delta Q/\Delta L$. But these ratios are \textit{not} the marginal
products. For instance, if labor is increased by this $\Delta L$, then an
additional $\Delta K$ must be used up to produce one more unit of output
($\Delta Q=1$) in a cost-minimizing manner. The usual ``marginal product'' of labor $MP_{L}$ is the extra product
produced per extra unit of labor if the production technique is hypothetically
shifted so that \textit{no more extra capital} is used, as illustrated in Figure \ref{fig:fig3-mpl}.

In our simple model, the marginal results of labor can be calculated by
dividing the marginal whole product through by $\Delta L$ to obtain $(1/\Delta
L,-\Delta K/\Delta L,-1)$. Since labor also creates the marginal unit of labor
$(0,0,1)$, the marginal whole product of labor is the following vector sum.

\begin{center}
	$\mathbf{MWP}_{\mathbf{L}}=(1/\Delta L,-\Delta K/\Delta L,0)=(1/\Delta
	L,-\Delta K/\Delta L,-1)+(0,0,1)$.
	
	Marginal Whole Product of Labor
\end{center}

\noindent Multiplying through by the prices yields the corresponding value
(using $\frac{p-MC}{\Delta L}=\frac{p-r\Delta K}{\Delta L}-\frac{w\Delta
	L}{\Delta L}$).

\begin{center}
	$\mathbf{P}\cdot\mathbf{MWP}_{\mathbf{L}}=(p,r,w)\cdot(1/\Delta L,-\Delta
	K/\Delta L,0)=(p-r\Delta K)/\Delta L=w+(p-MC)/\Delta L$.
	
	Value of Marginal Whole Product of Labor
\end{center}

If $\mathbf{P}\cdot\mathbf{MWP}_{\mathbf{L}}$ exceeds $w$ (the opportunity
cost of the marginal unit of labor), then it is profitable to increase the
amount of labor to produce more output by using up more capital services.
Conversely, if $\mathbf{P}\cdot\mathbf{MWP}_{\mathbf{L}}$ is less then $w$,
then the use of the marginal unit of labor does not cover its opportunity cost
so it would be better to reduce the amount of labor. Thus for profits to be
maximized, the value of the marginal whole product of labor must equal the
opportunity cost of labor.

\begin{center}
	$\mathbf{P}\cdot\mathbf{MWP}_{\mathbf{L}}=w$.
	
	Profit Max Implies: Value of Marginal Whole Product of Labor = Wage
\end{center}

\noindent Since $\mathbf{P}\cdot\mathbf{MWP}_{\mathbf{L}}=w+(p-MC)/\Delta L$,
the above result is equivalent to the usual $p=MC$.

\section{Comparison of the Two Treatments of MP Theory}

We have given an alternative treatment of MP theory using vector marginal
products. We have also used the juridical notion of the responsible factor
(here taken as labor) to organize the presentation. The crux of the two
presentations is in the two marginal conditions concerning labor:

\begin{center}
	Conventional labor equation:\qquad$p\cdot MP_{L}=w$
	
	Alternative labor equation:\qquad$\mathbf{P}\cdot\mathbf{MWP}_{\mathbf{L}}=w.$
\end{center}

\noindent In the conventional labor equation, $p$ and $MP_{L}$ (as well as
$w$) are scalars. In the alternative equation, $\mathbf{P}$ and $\mathbf{MWP}%
_{\mathbf{L}}$ are vectors (while $w$ remains a scalar). The conventional
interpretation of $MP_{L}$ pictures labor as producing marginal products
without using up any inputs (the miracle of ``virgin
birth'' or immaculate production of marginal product). The
marginal whole product of labor $\mathbf{MWP}_{\mathbf{L}}$ gives the picture
of the marginal effect of labor as producing outputs by using up other inputs
at minimum costs.

One could, of course, take capital services as the active or responsible
factor, define the marginal whole product of capital as $\mathbf{MWP}%
_{\mathbf{K}}=(1/\Delta K,0,-\Delta L/\Delta K)$, and then show that the
following condition is also equivalent to profit maximization (when costs are minimized).

\begin{center}
	$\mathbf{P}\cdot\mathbf{MWP}_{\mathbf{K}}=r$
	
	Profit Max Implies: Value of Marginal Whole Product of Capital = Rental
\end{center}

But instead of restoring a peaceful symmetry, this only highlights the
conflict since one cannot plausibly represent both capital as producing the
product by using labor, and labor as producing the product by using capital.
MP theory is not a juridical theory. It provides no grounds for choosing one
of the conflicting pictures over the other---for choosing the picture of the
burglar using tools to commit the crime over the picture of the tools using
the burglar to commit the crime. The distinction between the two pictures
comes from jurisprudence, not from economics. That is why such elementary
facts from jurisprudence need to be put (back) into economics
\cite{ellerman:juris-book}.

\section{Conclusion: The ``Advantages'' of
	Scalar MP Theory}

There are some basic problems with the usual scalar MP theory story. The
``miracle'' problem is that no factor $x_{i}$
can produce a part of the product without using up other factors, and the
usual ``story'' only seems that way by
assuming a conjectural shift to a more $x_{i}$-intensive production process so
that $MP_{i}=\partial y/\partial x_{i}$ more units are produced using the
extra $x_{i}$ and the same total amounts of the other factors. But that is
only a fictitious story since the neoclassical firm would only expand output
along the least-cost expansion path which would involve using up increments in
the other factors along with the increase in some input $x_{i}$. Hence the
non-fictitious notion of the marginal product of an additional unit of the
$x_{i}$ is the \textit{vectorial }marginal product of the factor that computes
all the marginal changes along the least-cost expansion path. 

Why doesn't the
economics profession use the mathematically equivalent theory with vector
marginal products describing motion along the least-cost expansion path? But
then the \textit{whole framing about distributive shares} in the outputs
collapses since it ignores the liabilities also created in production and it
ignores the actual appropriation of 100\% of the new property assets and
liabilities (the production vector or whole product) going to one party in a
productive opportunity.\footnote{In spite of obscene distribution of income
	and wealth resulting from centuries of leveraging or renting human beings,
	even the most progressive of neoclassical economists (e.g., Stiglitz, Krugman,
	Piketty, Sen, Bowles, Phelps, etc.) continue to frame the question in terms of the
	distributive shares picture--while ignoring the prior predistributive question
	of who is to ``be the firm'' (i.e., who is to
	appropriate the whole product) in the first place.} Thus it seems the
fictitious story about distributive shares based on scalar marginal product
theory was not a ``bug'' but a
``feature'' of the theory.

That brings us to the ``metaphor'' problem in
the usual presentation of MP theory. As von Wieser put it:

\begin{quotation}
	\noindent no one but the labourer could be named. Land and capital have no
	merit that they bring forth fruit; they are dead tools in the hand of man; and
	the man is responsible for the use he makes of them. \cite[p. 79]%
	{wieser:value}
\end{quotation}

\noindent Only the actions of the persons involved in a production opportunity
can be de facto responsible for the results, and the results are algebraically
symmetric, i.e., both negative and positive. In general terms, Labor $L$
(human actions of all the people working in a firm) uses up the services of
capital $K$ (including land) to produce the outputs $Q$. And since those human
actions are usually represented as an `input' $L$, Labor also produces $L$ and
uses up $L$ in the productive process. Hence the ``whole'' (both positive and negative) product of the
responsible factor $L$ is the whole product of labor vector $\mathbf{WP}%
_{\mathbf{L}}=\left(  Q,-K,0\right)  $ which can be represented as the whole
product $\left(  Q,-K,-L\right)  $ plus the labor services $\left(
0,0,L\right)  $. Hence the real Adding Up Theorem integrates the marginal
whole product of the responsible factor $L$ to see what that factor is
responsible for \textit{in} \textit{toto}. And by the fundamental theorem of
the calculus, the integration of $\mathbf{MWP}_{\mathbf{L}}$ is equivalent to
computing the net difference between when the people working in the productive
process carry out the actions $L$ and when they do nothing: $\left(
Q,-K,0\right)  -\left(  0,0,0\right)  =\mathbf{WP}_{\mathbf{L}}$.\footnote{In
	the Appendix, this computation is carried out for a Cobb-Douglas production
	function.}

The bigger ``problem'' with using vectorial
marginal products and using the standard juridical fact that only persons can
be responsible for anything is that it does not give a satisfactory `account'
of the standard employment system where the people working in the productive
process are rented,\footnote{``Since slavery was abolished,
	human earning power is forbidden by law to be capitalized. A man is not even
	free to sell himself: he must \textit{rent} himself at a
	wage.'' \cite[p. 52 (Samuelson's italics)]%
	{samuelson:economics} ``The labour market trades a commodity
	called `hours of labour services'. The corresponding price is the hourly wage
	rate. Rather loosely, we sometimes call this the `price of labour'. Strictly
	speaking, the hourly wage is the rental payment that firms pay to hire an hour
	of labour. There is no asset price for the durable physical asset called a
	`worker' because modern societies do not allow slavery, the institution by
	which firms actually own workers.'' \cite[p. 201]
	{beggetal:economics}} hired, or employed. Under the employment system, the
employees are only recognized as owning their labor $\left(  0,0,L\right)  $
which is sold in the employer-employee relationship to the employer (usually a
corporation) who pays off that labor liability $-L$ as well as the liabilities
for the other inputs $-K$ and thus gets the ownership of the product $Q$ so in
vectorial terms the employer is recognized as owning/owing the vector of
assets and liabilities $\mathbf{WP}=\left(Q,-K,-L\right)  $. Hence the
people working in the opportunity (including managers) are responsible for producing

\begin{center}
	$\mathbf{WP}_{\mathbf{L}}=\left(  Q,-K,0\right)  =\left(  Q,-K,-L\right)
	+\left(  0,0,-L\right)  =\mathbf{WP}+\left(  0,0,L\right)  $
\end{center}

\noindent but they only get the ownership of their labor $\left(
0,0,L\right)  $. As John Bates Clark put it:

\begin{quotation}
	\noindent A plan of living that should force men to leave in their employer's
	hands anything that by right of creation is theirs, would be an institutional
	robbery---a legally established violation of the principle on which property
	is supposed to rest. \cite[p. 9]{clark:dist-wealth}
\end{quotation}

\noindent The institutional robbery is the difference between what
``by right of creation is theirs'' $\mathbf{WP}_{\mathbf{L}}$ and what they are recognized as owning $\left(
0,0,L\right)  $, namely

\begin{center}
	$\mathbf{WP}_{\mathbf{L}}-\left(  0,0,L\right)  =\left(  Q,-K,0\right)
	-\left(  0,0,L\right)  =\left(  Q,-K,-L\right)  =\mathbf{WP}$
\end{center}

\noindent which is the whole product $\mathbf{WP}$, i.e. what accrues to
``being the firm.'' And the value of the
``institutional robbery'' $\mathbf{WP}$ is
the profits:

\begin{center}
	$\mathbf{P}\cdot\mathbf{WP}=\left(  p,r,w\right)  \cdot\left(  Q,-K,-L\right)
	=pQ-rK-wL$.
\end{center}

\noindent But institutionally, Labor is robbed of ``being the
firm,'' i.e., being a democratic firm \cite{ell:dwof}.

These results are summarized in the following Table 1.

\begin{center}%
	\begin{tabular}
		[c]{|c|c|c|}\hline
		Property imputations & Property Assets \& liabilities & Net
		value\\\hline\hline
		Labor de facto responsible for & $(Q,-K,0)$ & Value-added\\
		whole product of labor $WP_{L}$ &  & $pQ-rK$\\\hline\hline
		Labor legally appropriates & $\left(  0,0,L\right)  $ & Labor costs\\
		labor commodity $L$ &  & $wL$\\\hline\hline
		Labor responsible for & $\left(  Q,-K,0\right)  $ & \\
		but does not appropriate & \underline{ $-\left(  0,0,L\right)  $} & Profits\\
		whole product $WP$ & $=\left(  Q,-K,-L\right)  $ & $pQ-rK-wL$\\\hline
	\end{tabular}

	Table 1: Property imputations in an employment firm
\end{center}

While in value terms, Labor is robbed of the profit; Labor
is institutionally robbed of being the firm, i.e., a democratic firm (where Labor is robbed of
being the firm even in the special case of zero profits). This is obviously an
unsatisfactory `scientific account' for the whole system of renting persons.
Hence the orthodox economics profession presents MP theory as a `scientific'
theory using the ``virgin birth'' scalar
notion of marginal product and uses the \textit{metaphorical} notion of all
the causally efficacious factors of production as being ``responsible'' for their share of the outputs--which then
seems to satisfy the \textit{metaphorical} imputation principle:\footnote{From the viewpoint of the labor theory of property, such imitation is the sincerest form of flattery.}

\begin{quotation}
	\noindent The basic postulate on which the argument rests is the ethical
	proposition that an individual deserves what is produced by the resources he
	owns. \cite[p. 199]{friedman:pricetheory}
	
	\noindent The analysis [of market competition] shows how, under the conditions
	necessary for its existence, this organization achieves efficiency in the
	utilization of resources and justice in the distribution of the total product,
	efficiency being defined by the ends chosen by individuals and justice by the
	principle of equality in relations of reciprocity, giving each the product
	contributed to the total by its own performance (``what a man
	soweth that shall he also reap''). \cite[p. 292]
	{knight:history}
\end{quotation}

\noindent Thus considering the alternative, the ideological advantages of the
scalar MP theory are quite clear.

\bigskip

\section{Declarations}

\begin{itemize}
\item Funding: No funds, grants, or other support was received.
\item Conflict of interest/Competing interests: The author has no relevant financial or non-financial interests to disclose.
\end{itemize}

\noindent

\bigskip
\begin{appendices}
\section{Mathematical Appendix}

\subsection{Standard MP Theory}

Let $y=f(x_{1},...,x_{n})$ be a smooth neoclassical production function with
$p$ as the competitive unit price of the output $y$ and $w_{1},...,w_{n}$ as
the respective competitive unit prices of the inputs $x_{1},...,x_{n}$. The
cost minimization problem involves the input prices and a given level of
output $y_{0}$:

\begin{center}
	minimize: $C=\sum_{i=1}^{n}w_{i}x_{i}$
	
	subject to: $y_{0}=f\left(  x_{1},...,x_{n}\right)  $
	
	Minimize Cost to Produce Given Output
\end{center}

Forming the Lagrangian

\begin{center}
	$L=\sum_{i=1}^{n}w_{i}x_{i}-\lambda\left(  y_{0}-f\left(  x_{1},...,x_{n}%
	\right)  \right)  $,
\end{center}

the first-order conditions

\begin{center}
	$\frac{\partial L}{\partial x_{i}}=w_{i}-\lambda\frac{\partial f}{\partial
		x_{i}}=0$ for $i=1,...,n$
\end{center}

solve to: \qquad

\begin{center}
	$\lambda=\frac{w_{1}}{\partial f/\partial x_{1}}=...=\frac{w_{n}}{\partial
		f/\partial x_{n}}$.
	
	First-Order Conditions for Cost Minimization
\end{center}

These equations together with the production function determine the $n$
unknowns $x_{1},...,x_{n}$. Varying the input prices and level of output
parametrically determines the conditional factor demand functions:

\begin{center}
	$x_{1}=\varphi_{1}\left(  w_{1},...,w_{n},y\right)  $
	
	$\qquad\vdots$
	
	$x_{n}=\varphi_{n}\left(  w_{1},...,w_{n},y\right)  $.
	
	Conditional Factor Demand Functions
\end{center}

These functions give the optimum level of the inputs to minimize the cost to
produce the given level of output at the given input prices. Taking the input
prices as fixed parameters, we can write the conditional factor demand
functions as $x_{i}=\varphi_{i}(y)$ for $i=1,...,n$. These functions define
the cost-minimizing expansion path through input space parameterized by the
level of output. Substituting into the sum for total costs yields the

\begin{center}
	$C\left(  y\right)  =\sum_{i=1}^{n}w_{i}\varphi_{i}\left(  y\right)  $.
	
	Cost Function
\end{center}

\noindent Differentiation by $y$ yields the marginal cost function.

\qquad

\begin{center}
	$MC=\frac{dC}{dy}=\sum_{i=1}^{n}w_{i}\frac{\partial\varphi_{i}}{\partial y} $.
	
	Marginal Cost
\end{center}

\noindent The factor demand functions can also be substituted into the
production function to obtain the identity:

\begin{center}
	$y=f\left(  \varphi_{1}\left(  y\right)  ,...,\varphi_{n}\left(  y\right)
	\right)  $.
\end{center}

\noindent Differentiating both sides with respect to $y$ yields the useful equation:

\begin{center}
	$1=\sum_{i=1}^{n}\frac{\partial f}{\partial x_{i}}\frac{\partial\varphi_{i}%
	}{\partial y}$.
\end{center}

\noindent Multiplying both sides by the Lagrange multiplier allows us to
identify $\lambda$ as the marginal cost.

\begin{center}
	$\lambda=\sum_{i=1}^{n}\left(  \lambda\frac{\partial f}{\partial x_{i}%
	}\right)  \frac{\partial\varphi_{i}}{\partial y}=\sum_{i=1}^{n}w_{i}%
	\frac{\partial\varphi_{i}}{\partial y}=MC$.
	
	Lagrange Multiplier of Minimum Cost Problem is Marginal Cost
\end{center}

\noindent Using the customary marginal product notation, $MP_{i}=\partial
f/\partial x_{i}$ for $i=1,...,n$, the first order conditions for cost
minimization can be written as:

\begin{center}
	$MC=\frac{w_{1}}{MP_{1}}=...=\frac{w_{n}}{MP_{n}}$.
	
	Cost Minimization Conditions
\end{center}

The marginal products should not be confused with the reciprocals of the
factor demand function partials:

\begin{center}
	$\frac{\partial f}{\partial x_{i}}\neq1/\frac{\partial\varphi_{i}}{\partial
		y}$.
\end{center}

\noindent The marginal product $MP_{i}=\partial f/\partial x_{i}$ of $x_{i}$
gives the marginal increase in $y$ when there is both a marginal increase in
$x_{i}$ and a shift to a more $x_{i}$-intensive production technique so that
exactly the same amount of the other inputs is used. No factor prices or cost
minimization is involved in the definition. The reciprocal of $\partial
\varphi_{i}/\partial y$ gives the marginal increase in $y$ associated with a
marginal increase in $x_{i}$ when there is a corresponding increase in the
other inputs so as to produce the extra output at minimum cost.

\subsection{MP Theory with Product Vectors}

For the inclusive algebraically symmetric notion of the product, we will use
vectors with the outputs listed first followed by components for the inputs.
The positive product is $(y,0,...,0)$, the negative product is $(0,-x_{1}%
,...,-x_{n})$, and their sum is the

\begin{center}
	$\mathbf{WP}=(y,-x_{1},...,-x_{n})$
	
	\textit{Whole Product Vector}
\end{center}

The whole product vector is usually called the ``production
vector'' or ``net output
vector'' \cite[p. 8]{varian:micro} in the set-theoretic
presentations using production sets rather than production functions. Assuming
that costs are minimized at each output level, we can restrict attention to
the whole product vectors along the expansion path:

\begin{center}
	$\mathbf{WP}\left(  y\right)  =\left(  y,-\varphi_{1}\left(  y\right)
	,...,-\varphi_{n}\left(  y\right)  \right)  .$
\end{center}

The gradient $\triangledown_{y}=\frac{\partial}{\partial y}$ operator applied
to the whole product vector is the \textit{marginal whole product}
$\mathbf{MWP}$.

\begin{center}
	$\mathbf{MWP}\left(  y\right)  =\bigtriangledown_{y}\mathbf{WP}\left(
	y\right)  =\left(  1,-\frac{\partial\varphi_{1}}{\partial y},...,-\frac
	{\partial\varphi_{n}}{\partial y}\right)  $.
	
	Marginal Whole Product Vector $\mathbf{MWP}$
\end{center}

The price vector is $\mathbf{P}=(p,w_{1},...,w_{n})$, the value of the whole
product (the dot product of the price and whole product vectors) is the profit.

\begin{center}
	$\mathbf{P}\cdot\mathbf{WP}=py-\sum_{i=1}^{n}w_{i}x_{i}=py-C\left(  y\right)
	$,
	
	Value of Whole Product = Profit
\end{center}

\noindent and the \textit{value of the marginal whole product} is the

\begin{center}
	$\mathbf{P}\cdot\mathbf{MWP}(y)=\mathbf{P}\cdot\left(  1,-\frac{\partial
		\varphi_{1}}{\partial y},...,-\frac{\partial\varphi_{n}}{\partial y}\right)
	=p-\sum_{i=1}^{n}w_{i}\frac{\partial\varphi_{i}}{\partial y}=p-MC$.
	
	Value of Marginal Whole Product = Marginal Profit
\end{center}

The necessary condition for profit maximization is that the marginal whole
product has zero net value, which yields the familiar conditions $p=MC$.
Substituting $p$ for $MC$ in the cost minimization conditions yields the
central equations in the usual presentation of MP theory:

\begin{center}
	$pMP_{i}=w_{i}$ for $i=1,...,n$
\end{center}

\noindent which are interpreted as showing that in competitive equilibrium,
each unit of a factor is paid $w_{i}$ which is the value $pMP_{i}$ of
``what it produces'' $MP_{i}$.

\subsection{One Responsible Factor}

We move now to the formulation of the same mathematics but with certain
factors treated as responsible factors, i.e., the treatment of MP theory with
a responsible factor. We assume only one responsible factor that can be
arbitrarily taken as the first factor, which provides the services $x_{1}$. In
terms of totals, the responsible factor, by performing the services or actions
$x_{1}$, is responsible on the positive side for producing $y$ and is
responsible on the negative side for using up the other inputs $x_{2}%
,...x_{n}$. Since the customary notation lists $x_{1}$ along side the other
inputs, we could also picture the responsible factor as both producing and
using up $x_{1}$ (which thus cancels out). Thus the whole product of the
responsible factor is:

\begin{center}
	$\mathbf{WP}_{1}=\left(  y,0,-x_{2},...,-x_{n}\right)  =\mathbf{WP}+\left(
	0,x_{1},0,...,0\right)  $.
	
	Whole Product of Responsible Factor $x_{1}$
\end{center}

\noindent The whole product of the responsible factor is formally the sum of
the whole product and the services of the responsible factor.

Since we are now assuming only one responsible factor, we have the luxury of
mathematically treating its actions as the independent variable. Restricting
attention to the expansion path as usual and assuming $\partial\varphi
_{1}/\partial y\neq0$, we can invert the first factor demand function to obtain

\begin{center}
	$y=\varphi_{1}^{-1}\left(  x_{1}\right)  $
\end{center}

\noindent which can be substituted into the other factor demand functions to
obtain the other inputs as functions of $x_{1}$:

\begin{center}
	$x_{i}=\varphi_{i}\left(  \varphi_{1}^{-1}\left(  x_{1}\right)  \right)  $ for
	$i=2,...,n$.
\end{center}

\noindent The whole product of the responsible factor can then be expressed as
a function of $x_{1}$:

\begin{center}
	$\mathbf{WP}_{1}\left(  x_{1}\right)  =\left(  \varphi_{1}^{-1}\left(
	x_{1}\right)  ,0,-\varphi_{2}\left(  \varphi_{1}^{-1}\left(  x_{1}\right)
	\right)  ,...,-\varphi_{n}\left(  \varphi_{1}^{-1}\left(  x_{1}\right)
	\right)  \right)  $.
	
	Whole Product of Responsible Factor $x_{1}$ as a Function of $x_{1}$
\end{center}

We can now present a realistic picture of the effects of a marginal increase
in the responsible factor. A marginal increase in $x_{1}$ with both use up the
other factors at the rate

\begin{center}
	$\frac{\partial\varphi_{i}\left(  \varphi_{1}^{-1}\right)  }{\partial x_{1}%
	}=\frac{\partial\varphi_{i}/\partial y}{\partial\varphi_{1}/\partial y}$
\end{center}

\noindent and will increase the output at the rate

\begin{center}
	$\frac{\partial\varphi_{1}^{-1}}{\partial x_{1}}=\frac{1}{\partial\varphi
		_{1}/\partial y}$
\end{center}

\noindent along the expansion path. This information is given by the $x_{1}$
gradient $\triangledown_{1}=\frac{\partial}{\partial x_{1}}$ of the whole
product of the responsible factor, which is the marginal whole product of the
responsible factor:

\begin{center}
	$\mathbf{MWP}_{\mathbf{1}}=\bigtriangledown_{1}\mathbf{WP}_{\mathbf{1}}\left(
	x_{1}\right)  =\left(  \frac{1}{\partial\varphi_{1}/\partial y},0,\frac
	{\partial\varphi_{2}/\partial y}{\partial\varphi_{1}/\partial y}%
	,...,\frac{\partial\varphi_{n}/\partial y}{\partial\varphi_{1}/\partial
		y}\right)  $.
	
	Marginal Whole Product of Responsible Factor
\end{center}

This marginal whole product vector $\mathbf{MWP}_{\mathbf{1}}$ presents what
the responsible factor is marginally responsible for in quantity terms. Thus
it should be compared with the marginal product $MP_{1}$ in the conventional
treatment of MP theory. The marginal product $MP_{1}$ is fine as a
mathematically defined partial derivative. But to interpret it in terms of
production, one has to consider the purely notional shift to a more $x_{1}%
$-intensive productive technique so that exactly the same amount of the other
factors is consumed. That is \textit{not} how output changes in the
cost-minimizing firm. The marginal whole product $\mathbf{MWP}_{\mathbf{1}}$
presents the marginal changes in the output and the other factors associated
with a marginal increase in $x_{1}$ along the least-cost expansion path.

The value of the marginal whole product of $x_{1}$ is the dot product:

\begin{center}
	$\mathbf{P}\cdot\mathbf{MWP}_{\mathbf{1}}=\frac{\left[  p-w_{2}\partial
		\varphi_{2}/\partial y-...-w_{n}\partial\varphi_{n}/\partial y\right]
	}{\partial\varphi_{1}/\partial y}=\frac{\left[  p-MC\right]  }{\partial
		\varphi_{1}/\partial y}+w_{1}$.
	
	Value of Marginal Whole Product of Responsible Factor
\end{center}

Thus we have that the necessary condition for profit maximization, $p=MC$, is
equivalent to

\begin{center}
	$\mathbf{P}\cdot\mathbf{MWP}_{\mathbf{1}}=w_{1}$.
	
	Profit Max Implies Value of Marginal Whole Product
	
	= Opportunity cost of using marginal unit of responsible factor
\end{center}

Production is carried to the point where the value of the marginal whole
product of the responsible factor is equal to its opportunity cost given by
$w_{1}$. Since we are assuming cost minimization, this is also equivalent to
the conventional equation: $pMP_{1}=w_{1}$.

\subsection{Example: Cobb-Douglas Production Function}

For a Cobb-Douglas production function, $Q=AK^{a}L^{b}$ with the unit prices
of $K$ and $L$ being $r$ and $w$, one can derive the vectorial marginal
products by taking derivatives along the least cost expansion path. To fill in
the details, the first-order conditions for cost-minimization, $\frac{w_{1}%
}{\partial f/\partial x_{1}}=...=\frac{w_{n}}{\partial f/\partial x_{n}} $,
are in this case:

\begin{center}
	$\frac{r}{aAK^{a-1}L^{b}}=\frac{w}{bAK^{a}L^{b-1}}$ or $\frac{AK^{a}L^{b-1}%
	}{AK^{a-1}L^{b}}=\frac{K}{L}=\frac{aw}{br}$
\end{center}

\noindent so the conditions that hold along the least-cost expansion path are:

\begin{center}
	$\frac{K}{L}=\frac{aw}{br}$.
\end{center}

The whole product of the responsible factor $L$, previously $\mathbf{WP}%
_{1}=\left(  y,0,-x_{2},...,-x_{n}\right)  $, can then be stated directly as a
function of $L$ using $K=\frac{aw}{br}L$:

\begin{center}
	$\mathbf{WP}_{\mathbf{L}}=\left(  Q\left(  L\right)  ,-K\left(  L\right)
	,0\right)  =\left(  A\left(  \frac{awL}{br}\right)  ^{a}L^{b},-\frac{aw}%
	{br}L,0\right)  =\left(  A\left(  \frac{aw}{br}\right)  ^{a}L^{a+b},-\frac
	{aw}{br}L,0\right)  $.
\end{center}

\noindent Taking the derivative with respect to $L$ gives the marginal whole
product of the responsible factor labor:

\begin{center}
	$\mathbf{MWP}_{\mathbf{L}}=\left(  \left(  a+b\right)  A\left(  \frac{aw}%
	{br}\right)  ^{a}L^{a+b-1},-\frac{aw}{br},0\right)  $.
\end{center}

Since labor is the only responsible factor, one can compute its total
responsibility for the positive and negative results of production by
``adding up'' or integrating its marginal
whole product from $0$ to $L$ to obtain the result---which is the whole
product of the responsible factor: $(Q,-K,0)$:

\begin{center}
	$%
	{\textstyle\int\nolimits_{0}^{L}}
	\mathbf{MWP}_{\mathbf{L}}dl=%
	{\textstyle\int\nolimits_{0}^{L}}
	\left(  \left(  a+b\right)  A\left(  \frac{aw}{br}\right)  ^{a}l^{a+b-1}%
	,-\frac{aw}{br},0\right)  dl$
	
	$=\left(  \left.  A\left(  \frac{aw}{br}\right)  ^{a}l^{a+b}\right]  _{0}%
	^{L},\left.  -\frac{aw}{br}l\right]  _{0}^{L},0\right)  $
	
	$=\left(  A\left(  \frac{aw}{br}\right)  ^{a}L^{a+b},-\frac{aw}{br}L,0\right)
	$
	
	$=\left(  A\left(  \frac{K}{L}\right)  ^{a}L^{a+b},-\frac{K}{L}L,0\right)
	=\left(  AK^{a}L^{b},-K,0\right)  =\left(  Q,-K,0\right)  =\mathbf{WP}%
	_{\mathbf{L}}$.
\end{center}

Orthodox MP theory metaphorically represents each factor as producing a
certain share of the product $Q$, as if each input could produce some part of
the product without using up some of the other factors. Then under the
additional assumption of constant returns to scale ($a+b=1$ in this case), it
proves the ``Adding Up'' or ``Exhaustion of the Product Theorem'' that the shares add up to
the entire output $Q$.\footnote{See, for instance, Friedman \cite[p.
	194]{friedman:pricetheory}.} In the case at hand, the shares are:

\begin{center}
	$KMP_{K}=K\frac{\partial Q}{\partial K}=KAaK^{a-1}L^{b}=aQ$ and $LMP_{L}%
	=L\frac{\partial Q}{\partial L}=LbAK^{a}L^{b-1}=bQ$
\end{center}

\noindent so the sum of the ``shares'' is:

\begin{center}
	$KMP_{K}+LMP_{L}=\left(  a+b\right)  Q$
\end{center}

\noindent which equals $Q$ under the assumption of constant returns, $a+b=1$.

Continuing with the Cobb-Douglas example, we will have at profit maximization
both $pMP_{L}=w$ and $\mathbf{P}\cdot\mathbf{MWP}_{\mathbf{L}}=w$ so we might
compute both functions on the left hand side to compare them. The scalar MP of
labor is $MP_{L}=bQ/L$ so as a function of $L$,

\begin{center}
	$pMP_{L}=pbA\left(  \frac{aw}{br}\right)  ^{a}L^{a+b-1}$.
\end{center}

In the vectorial case, we have:

\begin{center}
	$\mathbf{P}\cdot\mathbf{MWP}_{\mathbf{L}}=p\left(  a+b\right)  A\left(
	\frac{aw}{br}\right)  ^{a}L^{a+b-1}-r\left(  \frac{aw}{br}\right)  $
	
	$=\frac{a+b}{b}\left[  pbA\left(  \frac{aw}{br}\right)  ^{a}L^{a+b-1}\right]
	-\frac{aw}{b}=\frac{a+b}{b}pbA\left(  \frac{K}{L}\right)  ^{a}L^{a+b-1}%
	-\frac{aw}{b}$
	
	$=\frac{a+b}{b}pMP_{L}-\frac{aw}{b}=\frac{a}{b}\left[  pMP_{L}-w\right]
	+pMP_{L}$.
\end{center}

\noindent Thus we finally have:

\begin{center}
	$\mathbf{P}\cdot\mathbf{MWP}_{\mathbf{L}}=\frac{a}{b}\left[  pMP_{L}-w\right]
	+pMP_{L}$
\end{center}

\noindent and solving for $pMP_{L}$, we have:

\begin{center}
	$pMP_{L}=\frac{b}{a+b}\left[  \mathbf{P}\cdot\mathbf{MWP}_{\mathbf{L}}%
	+\frac{aw}{b}\right]  =\frac{b\mathbf{P}\cdot\mathbf{MWP}_{\mathbf{L}}%
		+aw}{a+b}$
\end{center}

\noindent so the two functions are not the same. But when $pMP_{L}=w$, then
$\mathbf{P}\cdot\mathbf{MWP}_{\mathbf{L}}=w$ and when $\mathbf{P}%
\cdot\mathbf{MWP}_{\mathbf{L}}=w$, then $pMP_{L}=w$, so the two conditions,
one using the `virgin-birth' $MP_{L}$ and the other using the vectorial
$\mathbf{MWP}_{\mathbf{L}}$, are equivalent. This illustrates how MP theory
could just as well be presented using the vectorial marginal products.

\subsection{MP Theory Without Substitution}

There is a school of heterodox economics (the neo-Ricardian school associated
with Sraffa and to some extent Marx) that dismisses marginal productivity theory by assuming fixed
proportions and thus disputing the assumption of substitutability. 

We have criticized the usual interpretation of $MP_{i}$ as the
``product of the marginal unit of $x_{i}$'' on a number of grounds. For instance, a marginal increase in $x_{i}$ cannot
produce an increase in the output out of thin air. Other inputs will be
needed. The definition of the partial derivative $MP_{i}$ however assumes
substitutability in the sense that there is a shift to a slightly more $x_{i}$
intensive productive technique so that more output can be produced using
exactly the same amount of the other factors. Yet we have shown that such an
imaginary shift is not necessary to interpret marginal productivity theory. By
using vectorial notions of the product, the theory can be expressed quite
plausibly using marginal whole products computed along the cost-minimizing
expansion path.

The luxury of the alternative treatment of MP theory becomes a necessity when
there is no substitutability as in a Leontief or Sraffa input-output model
\cite{schefold:sraffa}. Hence we will give the alternative treatment of MP
theory in such a model.

We will consider an example where there are $n$ commodities $x_{1},...,x_{n}$
and labor $L$ where the latter is taken as the services of the responsible
factor. The technology is specified by the $n\times n$ matrix $\mathbf{A}%
=[a_{ij}]$ where $a_{ij}$ gives the number of units of the $i^{th}$ good
needed per unit of the $j^{th}$ good as output. Thus for the output column
vector $\mathbf{x}=(x_{1},...,x_{n})^{T}$ (the superscript "$T$" denotes the
transpose), the vector of required commodity inputs is $\mathbf{Ax}$. The
labor requirements per unit are given by the vector $\mathbf{a}_{0}%
=(a_{01},...,a_{0n})$, so the total labor requirement is the scalar
$L=\mathbf{a}_{0}\mathbf{x}$.

Let $\mathbf{p}=(p_{1},...,p_{n})$ be the price vector and let $w$ be the wage
rate. We assume that the outputs and inputs are separated by one time period
(a ``year'') and that $r$ is the annual
interest rate. The competitive equilibrium condition is usually stated as the
zero-profits condition with no mention of marginal productivity or the like.
With labor taking its income at the end of the year, the zero-profit condition
for any output vector is:

\begin{center}
	$\mathbf{px}=(1+r)\mathbf{pAx}+w\mathbf{a}_{0}\mathbf{x}$.
\end{center}

Since this must hold for any $\mathbf{x}$, we can extract the following vector
equation \cite[p. 12]{schefold:sraffa}.

\begin{center}
	$\mathbf{p}=(1+r)\mathbf{pA}+w\mathbf{a}_{0}$
	
	Competitive Equilibrium Condition
\end{center}

We now show how this condition can be derived using MP-style reasoning with
products represented as vectors. The whole product will be a $2n+1$ component
column vector since the output vector $\mathbf{x}$ is produced a year after
the input vector $\mathbf{Ax}$. The following notation for the whole product
is self-explanatory:

\begin{center}
	$\mathbf{WP}=%
	\begin{bmatrix}
		\mathbf{x}\\
		-\mathbf{Ax}\\
		-\mathbf{a}_{0}\mathbf{x}%
	\end{bmatrix}
	$
	
	Whole Product Vector $\mathbf{WP}$
\end{center}

The whole product of the responsible factor is, as always, the sum of the
whole product and the services of the responsible factor (since the factor is
represented as both producing and using up its own services):

\begin{center}
	$\mathbf{WP}_{L}=\mathbf{WP}+%
	\begin{bmatrix}
		0\\
		0\\
		\mathbf{a}_{0}\mathbf{x}%
	\end{bmatrix}
	=%
	\begin{bmatrix}
		\mathbf{x}\\
		-\mathbf{Ax}\\
		0
	\end{bmatrix}
	$
	
	Whole Product of Labor Vector $\mathbf{WP}_{\mathbf{L}}$
\end{center}

To consider output variations, we use the output unit vectors $\mathbf{\delta
}_{j}=(0,...,0,1,0,...,0)^{T}$ where the $1$ is in the $j^{th}$ place. The
marginal whole product of the responsible factor with respect to the $j^{th}$
output is will be symbolized as:

\begin{center}
	$\mathbf{\triangledown}_{j}\mathbf{WP}_{L}=%
	\begin{bmatrix}
		\mathbf{\delta}_{j}\\
		-\mathbf{A\delta}_{j}\\
		0
	\end{bmatrix}
	$
	
	Marginal Whole Product of Labor with Respect to the $j^{th}$ Output
\end{center}

\noindent and the required labor is $\mathbf{a}_{0}\mathbf{\delta}_{j}=a_{0j}$
with the opportunity cost of $wa_{0j}$. The price vector stated in year-end
values is $\mathbf{P}=(\mathbf{p},(1+r)\mathbf{p},w$) so the value of the
marginal whole product of labor is:

\begin{center}
	$\mathbf{P}\cdot\mathbf{\triangledown}_{j}\mathbf{WP}_{L}=p_{j}-\left(
	1+r\right)  \mathbf{pA\delta}_{j}$ 
	
	Value of Marginal Whole Product of Labor with respect to the $j^{th}$ Output
\end{center}

When the value of that marginal whole product of the responsible factor with
respect to the $j^{th}$ output is set equal to opportunity cost of the
necessary labor $wa_{0j}$ for $j=1,...,n$, then we again have the same
equilibrium conditions:

\begin{center}
	$\mathbf{p}-(1+r)\mathbf{pA}=w\mathbf{a}_{0}$.
	
	Competitive Equilibrium Condition Expressed as: Value of Marginal Whole
	Product of Labor with Respect to Each Output = Its Opportunity Cost.
\end{center}

Thus the alternative presentation of MP theory with product vectors and
responsible factors can be used in models without substitution where the
conventional marginal products are undefined. Substitutability is only needed
for the \textit{scalar} marginal products. The marginal productivity analysis
works fine in the Leontief-Sraffa model with fixed proportions using the
vectorial marginal product of labor. Sraffians and neo-Ricardians in general
need to cast a wider net, e.g., into the field of jurisprudence, in order to
capture a substantive critique of marginal productivity theory.

\end{appendices}


\begin{thebibliography}{99}                                                                                               %
	
	
	\bibitem {arrow:firmingetheory}Arrow, Kenneth J. 1971. The Firm in General
	Equilibrium Theory. In \textit{The Corporate Economy}, edited by R. Marris and
	A. Woods. Cambridge: Harvard University Press.
	
	\bibitem {arrow-debreu:gemodel}Arrow, Kenneth J., and Gerard Debreu. 1954.
	Existence of an Equilibrium for a Competitive Economy. \textit{Econometrica
	}22: 265--90.
	
	\bibitem {arrow-hahn:gca}Arrow, Kenneth J., and Frank H. Hahn. 1971.
	\textit{General Competitive Analysis}. Holden-Day.
	
	\bibitem {beggetal:economics}Begg, David, Stanley Fischer, and Rudiger
	Dornbusch. 1997. \textit{Economics (Fifth Ed.)}. McGraw-Hill Co.
	
	\bibitem {clark:dist-wealth}Clark, John Bates. 1899. \textit{The Distribution
		of Wealth}. New York: Macmillan.
	
	\bibitem {debreu:theoryofvalue}Debreu, Gerard 1959. \textit{Theory of Value.}
	John Wiley \& Sons.
	
	\bibitem {ell:dwof}Ellerman, David. 1990. \textit{The Democratic Worker-Owned
		Firm}. Unwin-Hyman Academic. https://www.ellerman.org/the-democratic-worker-owned-firm/.
	
	\bibitem {ellerman:pandc}Ellerman, David. 1992. \textit{Property \& Contract
		in Economics: The Case for Economic Democracy}. Blackwell.
	
	\bibitem {ellerman:intro-lpt}Ellerman, David. 2014. On Property
	Theory.\textit{\ Journal of Economic Issues} XLVIII (3 Sept.): 601--24.
	
	\bibitem {ellerman:challengelpt}Ellerman, David. 2017. On the Labor Theory of
	Property: Is the Problem Distribution or Predistribution? \textit{Challenge:
		The Magazine of Economic Affairs} 60 (2): 171--88.
	
	\bibitem {ellerman:juris-book}Ellerman, David. 2021. \textit{Putting
		Jurisprudence Back into Economics: On What Is Really Wrong with Today's
		Neoclassical Theory}. SpringerNature. https://doi.org/10.1007/978-3-030-76096-0.
	
	\bibitem {fairchild:appliedsoc}Fairchild, Henry Pratt. 1916. \textit{Outline
		of Applied Sociology}. Macmillan.
	
	\bibitem {foxwell:intro}Foxwell, H. S. 1899. Introduction. In \textit{The
		Right to the Whole Produce of Labour}, by Anton Menger, v-cx. MacMillan.
	
	\bibitem {friedman:pricetheory}Friedman, Milton. 1976. \textit{Price Theory}. Aldine.
	
	\bibitem {friedman:capandfreedom}Friedman, Milton. 2002. \textit{Capitalism
		and Freedom (Fortieth Anniversary Ed.)}. University of Chicago Press.
	
	\bibitem {haavelmo:capital}Haavelmo, Trygve. 1960. \textit{A Study in the
		Theory of Investment}. University of Chicago Press.
	
	\bibitem {hirsh:invest}Hirshleifer, Jack 1970. \textit{Investment, Interest,
		and Capital}. Prentice-Hall.
	
	\bibitem {hodgskin:property}Hodgskin, Thomas. 1973. \textit{The Natural and
		Artificial Right of Property Contrasted}. Augustus M. Kelley.
	
	\bibitem {kant:metamorals}Kant, Immanuel. 1965 (1797). \textit{The
		Metaphysical Elements of Justice: Part I of The Metaphysics of Morals}.
	Translated by John Ladd. Bobbs-Merrill.
	
	\bibitem {keen:debunk}Keen, Steve. 2011. \textit{Debunking Economics--Revised
		and Expanded Edition: The Naked Emperor Dethroned?}. Zed Books.
	
	\bibitem {knight:history}Knight, Frank. 1956. \textit{On the History and
		Method of Economics}. Phoenix Books.
	
	\bibitem {knight:risk}Knight, Frank.1965. \textit{Risk, Uncertainty and
		Profit}. Harper Torchbooks.
	
	\bibitem {menger:wholeproduct}Menger, Anton. 1899. \textit{The Right to the
		Whole Produce of Labour: The Origin and Development of the Theory of Labour's
		Claim to the Whole Product of Industry}. Trans. M. E. Tanner. Macmillan and Co.
	
	\bibitem {milljames:elements}Mill, James. 1826. \textit{Elements of Political
		Economy}. 3rd ed. London.
	
	\bibitem {quirk-saposnik}Quirk, J., and R. Saposnik. 1968.
	\textit{Introduction to General Equilibrium Theory and Welfare Economics}. McGraw-Hill.
	
	\bibitem {rawls:justice}Rawls, John. 1999. \textit{A Theory of Justice
		(Revised Ed.)}. Belknap Press.
	
	\bibitem {samuelson:economics}Samuelson, Paul A. 1976. \textit{Economics (10th
		ed.),} McGraw-Hill.
	
	\bibitem {schefold:sraffa}Schefold, Bertram. 1989. \textit{Mr Sraffa on Joint
		Production and Other Essays}. Unwin-Hyman Academic.
	
	\bibitem {theo:briggs}Theobald, Mark. 2004. \textit{Briggs Mfg. Co.,}
	Coachbuilt. http://www.coachbuilt.com/bui/b/briggs/briggs.htm.
	
	\bibitem {thurow:inequality}Thurow, Lester. 1975. \textit{Generating
		Inequality: Mechanisms of Distribution in the U.S. Economy}. Basic Books.
	
	\bibitem {varian:micro}Varian, Hal. 1984. \textit{Microeconomic Analysis. 2nd
		ed.} W.W. Norton.
	
	\bibitem {wieser:social}Wieser, Friedrich von. 1927. \textit{Social
		Economics}. Translated by A. Ford Hinrichs. Adelphi.
	
	\bibitem {wieser:value}Wieser, Friedrich von. 1930 (1889). \textit{Natural
		Value}. Trans. C. A. Malloch. G.E. Stechert and Company.
\end{thebibliography}
\end{document}